\documentclass[conference]{IEEEtran}
\IEEEoverridecommandlockouts

\usepackage{cite}
\usepackage{amsmath,amssymb,amsfonts}
\usepackage{algorithmic}
\usepackage{graphicx}
\usepackage{textcomp}
\usepackage{xcolor}
\usepackage{float}
\usepackage{url}

\def\BibTeX{{\rm B\kern-.05em{\sc i\kern-.025em b}\kern-.08em
    T\kern-.1667em\lower.7ex\hbox{E}\kern-.125emX}}

\begin{document}

\title{Implementing Transport Coding in OMNeT++ for Message Delay Reduction}

\author{
\IEEEauthorblockN{Ilya Petrovanov}
\IEEEauthorblockA{\textit{HSE University, Moscow Institute of} \\
\textit{Electronics and Mathematics}\\
Moscow, Russia \\
ipetrovanov@hse.ru}
\and
\IEEEauthorblockN{Anton Sergeev}
\IEEEauthorblockA{\textit{HSE University, Moscow Institute of} \\
\textit{Electronics and Mathematics}\\
Moscow, Russia \\
avsergeev@hse.ru}
}

\maketitle

\begin{abstract}
Transport coding reduces message delay in packet-switched networks by introducing controlled redundancy at the transport layer: $k$ original packets are encoded into $n\ge k$ coded packets, and the message is reconstructed after the first $k$ successful deliveries, effectively shifting latency from the maximum packet delay to the $k$-th order statistic. We present a concise, reproducible discrete-event implementation of transport coding in OMNeT++, including a multi-hop Kleinrock-type network, FIFO queues, exponential service and link delays, and explicit receiver-side reconstruction that records message delay and deadline violations. Using paired uncoded ($n{=}k$) and coded ($n{>}k$) configurations at the same message generation rate, we compare delay, reliability, and saturation effects across code rates and input loads. Simulation results show consistent reductions of average delay and late-delivery probability for moderate redundancy, while keeping the saturation throughput close to the uncoded baseline. The proposed model provides a transparent bridge between analytical transport-coding formulas and executable simulation for tuning redundancy in low-latency services.

\end{abstract}


\section{Introduction}
Many latency-sensitive services require not only small average delay but also a low probability of exceeding a strict deadline $t_{\max}$. In packet-switched networks, message delivery time is often dominated by the slowest packet, which increases both mean delay and tail latency under load.

Transport coding addresses this by adding controlled redundancy at the transport layer: a message is split into $k$ packets, encoded into $n\ge k$ packets, and decoded once any $k$ packets arrive. Analytically, this replaces the ``last-packet'' delay metric with the $k$-th order statistic, which can reduce both delay and variance. While the theoretical effect is well understood, its end-to-end behavior in a concrete multi-hop network with queues and reconstruction logic is less documented.

This paper contributes a minimal yet explicit OMNeT++ discrete-event model of transport coding, including FIFO queueing, exponential service and link delays, and message-level reconstruction with deadline accounting. We use paired uncoded ($n{=}k$) and coded ($n{>}k$) configurations at the same message rate to quantify delay, tail behavior, and saturation. The resulting framework serves as a bridge between analytical formulas and reproducible simulation code, enabling practical tuning of redundancy for low-latency networks.

\section{Analytical Foundations of Transport Coding}
Transport coding introduces redundancy at the transport layer: a message of $k$ packets is encoded into $n\ge k$ packets (e.g., via an MDS code), and decoding succeeds after any $k$ packets arrive. Hence, the message latency changes from a ``last-packet'' event to the $k$-th successful arrival among $n$ transmissions.

We follow a Kleinrock-type model with Poisson packet arrivals of rate $\lambda$, identical links of capacity $c$, and service parameter $\mu$. The channel load is
\begin{equation}
\rho=\frac{\lambda}{\mu c}.
\end{equation}
Let the per-packet delay be exponential with mean
\begin{equation}
t(\rho)=\frac{\lambda}{\mu c \gamma}\cdot\frac{1}{1-\rho},
\end{equation}
where $\gamma$ is a normalization factor. For an \emph{uncoded} $k$-packet message, the delay equals the maximum among $k$ i.i.d. packet delays:
\begin{equation}
T_{\text{unc}}=\mathbb{E}[T_{k:k}]=t(\rho)\sum_{i=1}^{k}\frac{1}{i}.
\end{equation}

With coding, the rate is $R=k/n$ and the effective load increases to $\rho/R$, giving
\begin{equation}
t\!\left(\frac{\rho}{R}\right)=\frac{\lambda}{\mu c \gamma}\cdot\frac{R}{R-\rho},
\qquad
T_{\text{cod}}=\mathbb{E}[T_{k:n}]=t\!\left(\frac{\rho}{R}\right)\sum_{i=n-k+1}^{n}\frac{1}{i}.
\end{equation}
The relative gain is then
\begin{equation}
f(R)=\frac{T_{\text{unc}}}{T_{\text{cod}}}
=\frac{1-\rho/R}{1-\rho}\cdot
\frac{\sum_{i=1}^{k}\frac{1}{i}}{\sum_{i=n-k+1}^{n}\frac{1}{i}}.
\end{equation}
This expression captures the main trade-off: redundancy reduces order-statistics latency but increases load via $\rho/R$, so the best performance is typically achieved at moderate redundancy.

\section{Simulation Model}
We implement the proposed discrete-event model in \textbf{OMNeT++}. The network is a multi-hop non-regular grid with multiple alternative paths between a designated source and a receiver placed in opposite corners.

\textbf{Topology and delays.} Each bidirectional link has capacity $C=10$~Mbps and an exponential transmission/propagation delay with mean $d=2$~ms. Each node maintains a single FIFO queue and serves packets with exponential service time of mean $0.1$~ms. We assume error-free links and no packet losses.

\textbf{Traffic and coding.} Traffic is generated as messages parameterized by $(k,n)$: $k$ is the number of information packets and $n\ge k$ is the number of transmitted packets. We compare (i) uncoded baseline ($n=k$) and (ii) coded transmission ($n>k$) under the same message-generation process; the offered load is controlled by varying the inter-message interval, while redundancy is controlled via the code rate $R=k/n$.

\textbf{Forwarding and reconstruction.} Upon arrival, packets join the FIFO queue; after service completion, a packet is forwarded by selecting an outgoing link uniformly at random, producing diverse multi-hop routes. The receiver tracks each message by its identifier and declares completion upon the arrival of the first $k$ packets; the message delay is recorded as the time from creation to the latest arrival among these $k$ packets. We additionally record whether the delay exceeds a deadline $t_{\max}$ and compute delivered throughput.
\section{Results of Simulation}

We evaluate the model by comparing, for each load level, an uncoded baseline ($n{=}k$) and a coded configuration ($n{>}k$) with the same $k$ and the same message-generation process, so that only redundancy changes the packet-level load.

\textbf{Throughput.} As the offered load increases, the delivered throughput grows close to linearly and then saturates. The maximum delivered throughput in coded and uncoded scenarios differs only by a few percent; both approach approximately $9$--$9.5$~Mbps in the considered multi-hop topology with $10$~Mbps links.

\textbf{Mean delay and variability.} In the medium-to-high load region, coding yields a stable reduction of average message delay by about $9$--$11\%$ (e.g., from $\approx 0.26$--$0.34$~s without coding to $\approx 0.23$--$0.30$~s with coding). The delay distribution becomes more concentrated, and the variance of message delay decreases by roughly $25$--$35\%$, indicating reduced tail latency.

\textbf{Deadline reliability.} For the deadline $t_{\max}=0.3$~s, the probability of deadline violation in the most loaded regime decreases by more than a factor of two, from about $0.35$ to about $0.15$.

\textbf{Redundancy trade-off.} The empirical gain $f(R)=\text{delay}_{\text{uncoded}}/\text{delay}_{\text{coded}}$ exceeds $1$ over a broad range of code rates $R=k/n$. The best trade-off is achieved with moderate redundancy: $R\approx 0.5$--$0.8$ provides delay gains of about $1.05$--$1.1$, while excessive redundancy yields diminishing returns due to increased network load.




\section{Discussion and Conclusions}

Transport coding provides a practical latency--reliability trade-off in multi-hop packet-switched networks: moderate redundancy reduces average delay, suppresses long-delay tails, and lowers the probability of missing a deadline, while leaving the saturation throughput close to the uncoded baseline. At the same time, excessive redundancy increases the offered load and leads to diminishing returns, so code-rate tuning is essential.

Overall, the presented OMNeT++ model bridges analytical transport-coding intuition based on order statistics with an explicit discrete-event implementation that measures message-level delay and deadline violations. The framework can be reused to explore parameter settings $(k,n)$ for low-latency services and to extend experiments toward more realistic routing, traffic mixes, and queueing disciplines.

\section{Acknowledgement}
This work is an output of a research project implemented as part of the Basic Research Program and at the National Research University Higher School of Economics (HSE University).

This research was supported in part through computational resources of HPC facilities at the National Research University Higher School of Economics (HSE University).


\end{document}